\documentclass[prl, showpacs, twocolumn, floatfix]{revtex4}
\usepackage{latexsym}
\usepackage{amsmath}
\usepackage{amsfonts}
\usepackage{graphicx}
\usepackage{exscale}
\usepackage{amssymb}
\usepackage{mathrsfs}

\newcommand{\mf}[1]{\boldsymbol{#1}}

\begin{document}
\title{Phase-matched coherent hard x-rays from relativistic high-order harmonic generation}

\author{Markus C. Kohler, Michael Klaiber}
\affiliation{ Max-Planck-Institut f\"ur Kernphysik, Saupfercheckweg 1, D-69117 Heidelberg, Germany}
\author{Karen Z. Hatsagortsyan}\thanks{\mbox{Corresponding author: k.hatsagortsyan@mpi-k.de }}
 \affiliation{
Max-Planck-Institut f\"ur Kernphysik, Saupfercheckweg 1, D-69117
Heidelberg, Germany}
\author{Christoph H. Keitel}
\affiliation{ Max-Planck-Institut f\"ur Kernphysik, Saupfercheckweg 1, D-69117 Heidelberg, Germany}

\date{\today}

\begin{abstract}

High-order harmonic generation (HHG) with relativistically strong
laser pulses is considered employing electron ionization-recollisions from
multiply charged ions in counterpropagating, linearly polarized attosecond
pulse trains. The propagation of the harmonics through the medium and the scaling of HHG  into the
multi-kilo-electronvolt regime are investigated. 
We show that the phase mismatch caused by the free electron background can be compensated by an additional phase of the emitted harmonics specific to the considered setup which depends on the delay time between
the pulse trains. This renders feasible the phase-matched emission of harmonics with photon energies of
several tens of kilo-electronvolt from an underdense plasma.

\end{abstract}

\pacs{42.65.Ky, 42.79.Nv}

\maketitle

In the last decades, atomic high-order harmonic generation (HHG) in the non-relativistic regime
\cite{corkum} has been developed to a reliable source of coherent extreme ultraviolet (XUV)
radiation and of attosecond pulses opening the door for attosecond time-resolved
spectroscopy \cite{Krausz}. The further advancement of this technique into the hard x-ray domain would,
in particular, allow for ultrafast coherent diffraction imaging of single particles, clusters and biomolecules with sub-\aa ngstr\"om resolution,
tracking the electron motion in atoms and even for the investigation of time-resolved dynamics of nuclear excitations.
The large scale x-ray free electron lasers are likely to fulfill this task partly but are limited to energies around 10 keV.

Is it possible to extend the table-top HHG sources into the hard x-ray domain?
In principle, the harmonic photon energy can be increased by using stronger driving laser fields.
The state-of-the-art technique allows to generate coherent x-ray photons up to the keV energy range  \cite{seres} and to produce short XUV pulses of less than 100 as \cite{goulielmakis} from non-relativistic HHG in an atomic gas medium.
However, progress in this field appears to have reached a limit. Most importantly,
the further increase of the driving field intensity transfers the interaction
regime into the relativistic domain where the drift motion of the ionized electron
in the laser field propagation direction prohibits the recollision and,
consequently, suppresses HHG \cite{review}. This happens
at laser intensities above of $4\times10^{16}$ W/cm$^2$  at
infrared wavelengths, corresponding to the HHG cutoff
frequency of  $\omega_{c}\approx 10$ keV. This indicates the limit
of non-relativistic HHG. The second point hindering HHG at high
intensities is the phase-matching problem. In strong laser fields,
outer-shell electrons are rapidly ionized  and produce a large
free electron background causing phase mismatch between the
driving laser wave and the emitted x-rays.

Various methods for counteracting the relativistic drift have been
proposed. However, no universal solution has been found, each
method has its drawbacks. To suppress the drift, highly charged
ions moving relativistically against the laser propagation
direction~\cite{ion} or a gas of positronium atoms~\cite{collider}
can be used. Different combinations of laser fields also have been
proposed for this purpose such as a tightly focused laser beam
\cite{Becker:OL}, two crossed laser beams with linear polarization
\cite{Kylstra,Joachain} or with equal-handed circular polarization
\cite{circular}. In the latter field configuration, the
relativistic drift is eliminated, however, in this scheme the
phase-matching is especially problematic to realize
\cite{Chengpu}.
We have shown in \cite{klaiber_tailored,Verschl,ufo} that strong
attosecond pulse trains (APTs) employed as a driving field for HHG
can be very useful to suppress the relativistic drift. However,
all these efforts have only addressed HHG of a single atom rather than coherent
emission from a macroscopic gas target.

\begin{figure}[t]
\begin{center}
 \includegraphics[width=0.3\textwidth]{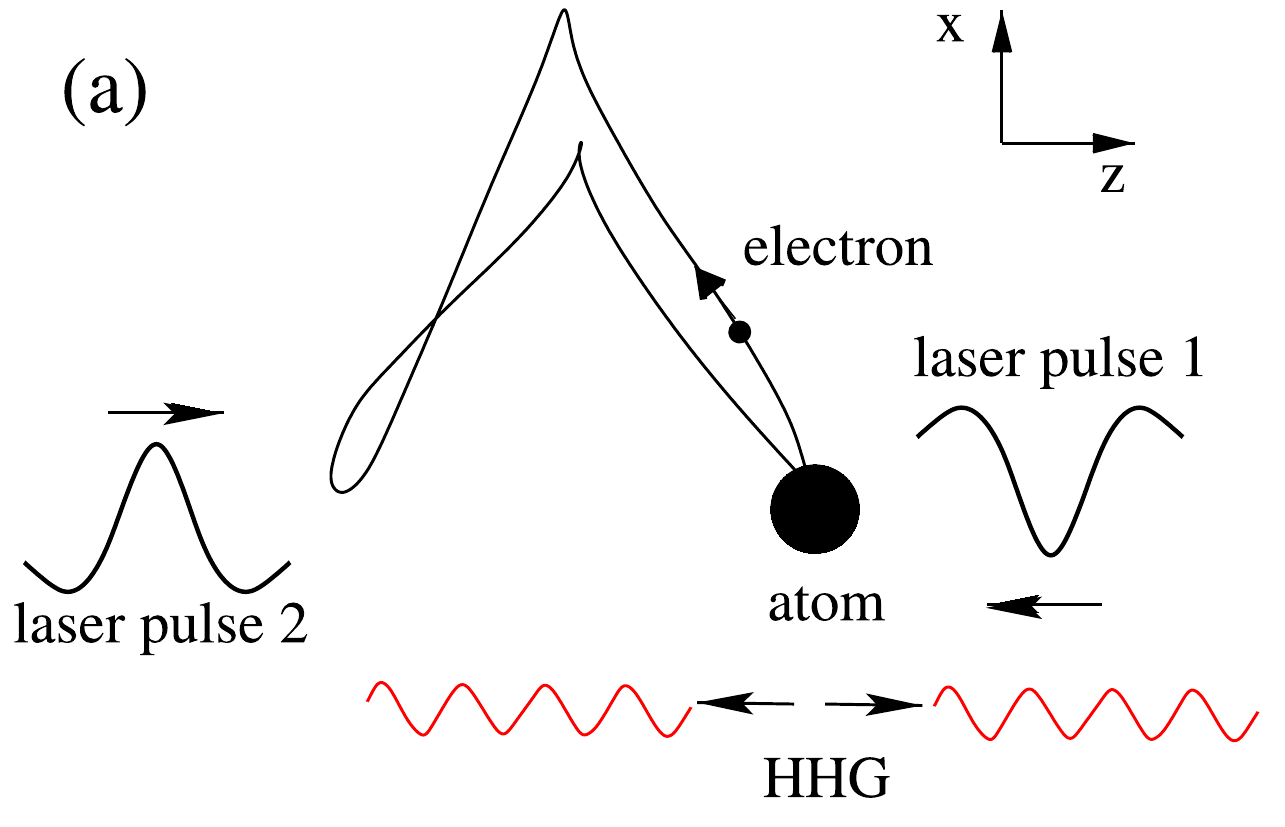}\\
\vskip 0.1cm
\includegraphics[width=0.49\textwidth]{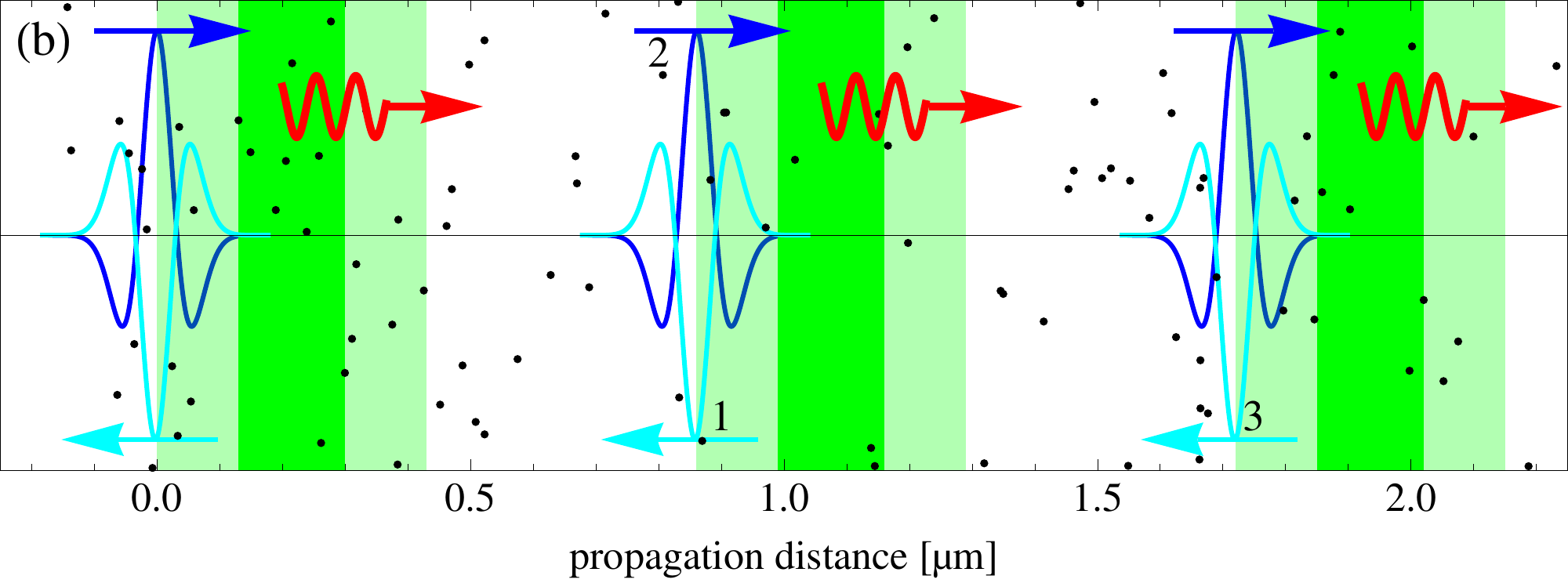}\\
\caption{(Color online) The HHG setup with counterpropagating APTs: (a) The trajectory of the rescattering electron;
(b) The contribution of different parts of the medium to HHG. The harmonics emitted from the green (dark) area propagates along the red (wavy) arrow. The HHG from shaded area is damped. The two driving APTs are shown in different blue color shades. The arrows indicate the propagation direction. } \label{setup}
\end{center}
\end{figure}

In this Letter, we investigate the feasibility of phase-matched harmonic emission from an
underdense plasma of multiply charged ions for a relativistic HHG setup employing two counterpropagating APTs.
We show that the HHG driven by counterpropagating APTs has an
additional intrinsic phase depending on the time delay between the pulses as well as on the pulse intensity. This phase avails to compensate the phase mismatch between the driving laser field and the emitted harmonics due to the free electron background. The latter can be achieved by modulating the driving field intensity with a slowly decreasing envelope. We have performed a complete, quantitative analysis of the macroscopic yield of the relativistic HHG
evidencing a small but detectable signal.

\begin{figure}[b]
\begin{center}
 \includegraphics[width=0.4\textwidth]{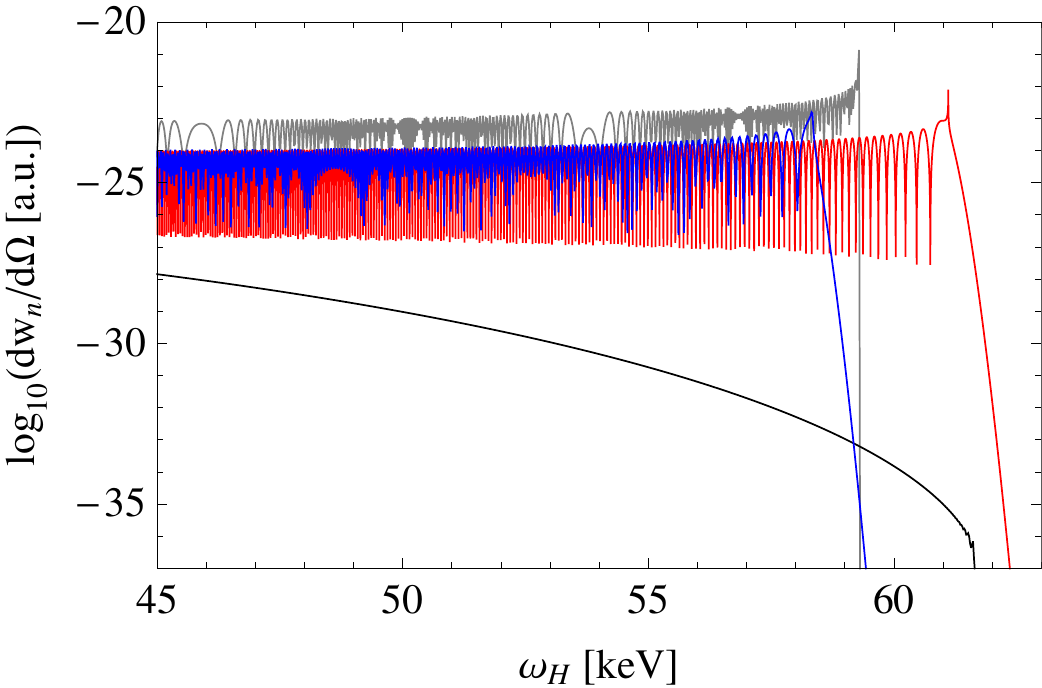}\\
\caption{(Color online) Single-atom HHG rates via \cite{ufo} in laser propagation direction:
(red, largest cutoff) for the discussed setup; the delay of pulses is $1.5$ fs, the laser field strength $E_0=21\text{ a.u.}$, and  $I_p=27.18$ (O$^{6+}$); (blue, lowest cutoff) for the same setup and parameters in DA including the mass shift; (black, bottom) for a conventional propagating laser field with $E_0=2.7\text{ a.u.}, \,I_p=7.35\text{ a.u.}$ (hydrogen-like ion),
and (gray, top) the latter within the DA including the mass shift. The indicated parameters are chosen such that both the cutoffs and the average ADK-tunneling rates are the same for the two fully-relativistic curves.
} \label{sa_comparison}
\end{center}
\end{figure}

The applied setup for relativistic HHG is shown in Fig.
\ref{setup}(a). The driving fields are two
counterpropagating APTs consisting of 100 as pulses with a peak intensity of the order of $10^{19}$ W/cm$^2$ and a spectral range of about 20eV.
Such pulses can be generated by employing the relativistic oscillating mirror of an overdense plasma surface in a strong laser field
\cite{Zepf}.
The electron is liberated by one pulse, it is driven in the
continuum and undergoes the relativistic drift. A
moment later, the second pulse reaches the electron, it reverts
the drift and realizes rescattering. The drift compensation is
very efficient as one can deduce from Fig. \ref{sa_comparison}
where the single-atom spectral emission rate in the present setup
is compared with that in the dipole approximation (DA). The setup
exhibits no significant drift any more. The rate for the applied
setup is much higher than the one for a conventional laser field
with the same cutoff. In the latter case, due to the drift, the rate would drop rapidly with increasing laser intensity.

We continue to estimate the macroscopic HHG yield of the given setup.
The contributions of different parts of the medium to the harmonic emission are shown in Fig. \ref{setup}(b).
Emission into the right direction is considered. HHG occurs only in small zones of the medium shown
as dark areas in Fig. \ref{setup}(b).
In this case, necessarily, recombination has to be arranged by a pulse propagating to the right. For instance, ions between pulse 1 and 3 in the sketch could have been ionized by pulse 1, which triggered the process.
In the next step, pulse 2 could arrange rescattering and HHG.
This is possible only up to a certain point (dark area)
since then pulse 2 and 3 meet and the process starts again. Contributions of atoms experiencing two pulses simultaneously (light shaded areas) are frustrated due to the chaotic trajectories of ionized electrons in this region \cite{chaotic}.
This limits the volume in longitudinal direction to about 1/3  and the possible delays are within $1 - 2$ fs.
After half a period, 1/3 of the white area would emit into the right direction.

In the following, we specify our model. As  HHG medium, an underdense plasma of O$^{6+}$
ions (ionization potential $I_p=27.18\text{ a.u.}$) is used which is
immediately formed when the first laser pulse of relativistic intensity
is applied to a  neutral atomic gas. This is because the outer shell
electrons of an oxygen atom are almost instantaneously ionized due to a much smaller binding
potential  (0.5 a.u. -- 5.1 a.u.) other than the two
remaining electrons in the closed $1s$-shell.  HHG is produced only by the
tightly-bound inner electron having an ionization potential
corresponding to the tunneling condition at relativistic
intensities. The O$^{6+}$ emission is slightly reduced by the
depletion to O$^{7+}$ whereas the O$^{7+}$ emission is not
phase-matched in the proposed phase-matching scheme. The driving
laser pulses are plane waves numerically propagated in the
relativistic free electron background using a
Crank-Nicolson-algorithm. The density of the free electrons is assumed
to be constant because the outer shell ionization time is small
compared to the laser period. Absorption of the high-frequency HHG
photons can be neglected because their energy is much higher than
the largest atomic transition energy. In order to find the overall
HHG yield, the photon spectral density $\frac{dN}{d\omega_H}$ in
the far field is calculated (in atomic units) \cite{Landau}:
\begin{equation}
 \frac{dN}{d\omega_H}=\frac{c}{4\pi^2\omega_H} R^2\int  d\Omega'  \vert\tilde{\mf{E}}(\mf{n}',\omega_H) \vert^2,
\end{equation}
where $\mf{n}'$ is the emission direction, $R$ the radius at an observation point, $\omega_H$ the positive harmonic frequency.
The spectral component of the electric field reads
\begin{equation}
 \tilde{\mf{E}}(\mf{n}',\omega_H)
= i \frac{\omega_H\rho e^{-i \omega_H R/c}}{Rc^2} \int d^3x_a  \tilde{\mf{j}_a}(\mf{x}_a,\omega_H,\mf{n}')\label{E_0},
\end{equation}
where $\rho$ is the density of the uniformly distributed ions, $\mf{j}_a(\mf{x},t,\mf{x}_a)$ the current density at a space-time point $\mf{x}$ and $t$ of a single ion located at $\mf{x}_a$ and $\tilde{\mf{j}_a}(\mf{x}_a,\omega_H,\mf{n}')$ its Fourier transform which is calculated quantum-mechanically using the relativistic strong field approximation (SFA) \cite{SFA}.
The emitted harmonics have the same polarization direction (along the x-axis) as the incident laser field.
The x-component of the spectral electron current density at a single multiply charged ion
in the SFA  based on the Klein-Gordon equation reads \cite{milo_rel}:
\begin{eqnarray}
\tilde{j_a}(\mf{x}_a,\omega_H,\mf{n}')= \frac{1}{4c} \sqrt{\frac{\omega_H}{2\pi}} \int d^4 x\int d^4 x' \phi^*(\mf{x}-\mf{x}_a,t) \nonumber\\ \times V_H(x)G(x,x') \kappa(t')
V_{AI}(x')\phi(\mf{x}'-\mf{x}_a,t'), \label{j_a}
\end{eqnarray}
where $\phi (\mf{x}-\mf{x}_a)$ is the wave function of the bound
electron, $V_H(x)$ the interaction Hamiltonian of the electron
with the harmonic field, $V_{AI}(x)$ the potential of the ionic
core and $G(x,x')$ is the Green function describing the free
electron evolution in the counterpropagating laser fields defined
in \cite{ufo}. We included a tunneling correction factor
$\kappa(t)=\sqrt{w_{ADK}(t)/w_{K}(t)}$ upgrading the Keldysh
tunneling rate $w_{K}(t)$ comprised in the SFA to fit the
ADK-ionization rate \cite{Ivanov}. As discussed, we only consider
the relevant scenario of interaction where
the electron moves in different counterpropagating pulses
successively. Accordingly, in each stage of the excursion, we
approximate the Green function by the Volkov Green function in a
field of the appropriate single laser pulse (see \cite{ufo}).
Eq. \eqref{j_a} is evaluated in the saddle point approximation.

\begin{figure}[t]
\begin{center}
 \includegraphics[width=0.4\textwidth]{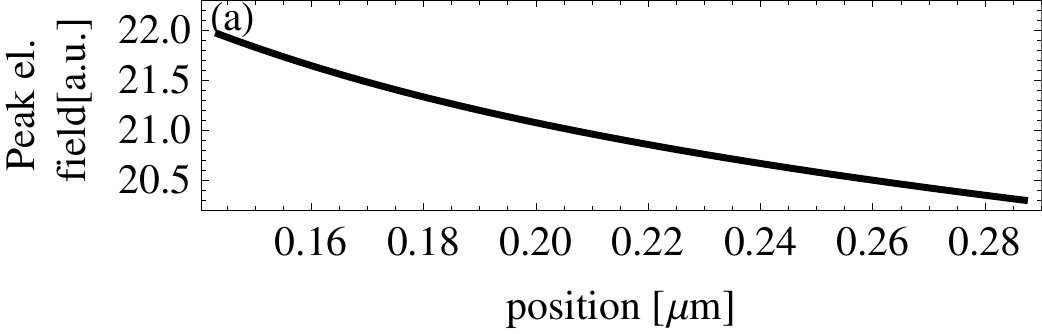}
 \includegraphics[width=0.4\textwidth]{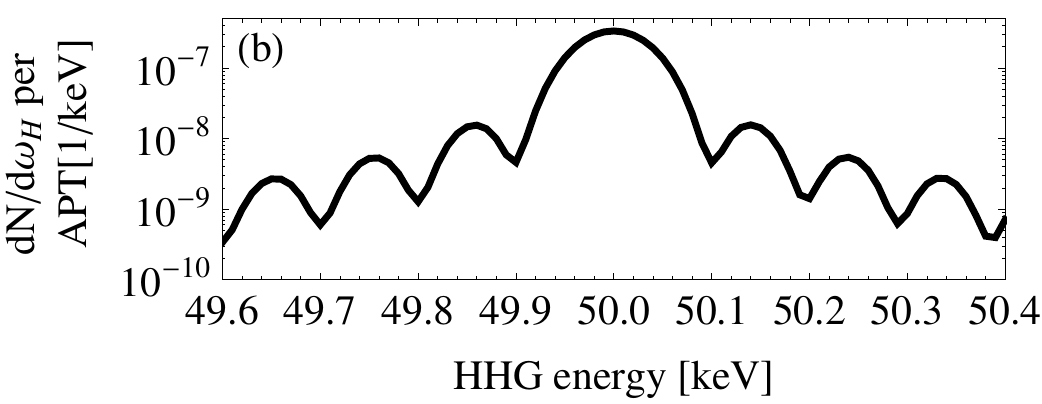}
\caption{(a) shows the optimized peak electric field variation in the first interaction zone. The excursion time increases to the right, (b)  Emitted spectral photon number (spikes at integer multiples of $\omega$ do not appear due to local averaging)  of the setup for the medium described in the text.} \label{figure2}
\end{center}
\end{figure}

Let us have a closer look at the phase difference of the harmonics emitted from different ions separated by a distance $\Delta z$ in the propagation direction:
\begin{equation}
 \Delta \varphi=\Delta \arg{\tilde{j_a}} \approx \Delta z\left(\frac{\omega_H}{c}\frac{\Delta v_g}{v_g} -\frac{\partial \varphi_{i}}{\partial z}\right)
\label{phase_delay}
\end{equation}
The first term describes the phase mismatch due to the free
electron dispersion with $\Delta v_g=\omega_p^2/2\omega_H^2$,
$\omega_p$ being the electron plasma frequency and $v_g$ the group
velocity of the driving laser pulse, whereas the last term is the
single-atom emission phase ($\varphi_i$)  depending on the laser
field conditions.
This intrinsic phase $\varphi_i$ is determined by the classical
action of the electron trajectory recolliding with the specific
harmonic energy and can be estimated as $\varphi_i\approx
\tilde{U}_{p}(\mf{r}_a)\tau(\mf{r}_a)$, with an effective
ponderomotive potential  $\tilde{U}_{p}(\mf{r}_a)$ and electron
excursion time $\tau(\mf{r}_a)$. Thus, $\varphi_i$ depends on the
laser intensity as well as on the delay between the two pulses.
The latter, being unique for this laser setup, mainly affects the
electron excursion time $\tau(\mf{r}_a)$ and varies along the
propagation direction. In order to achieve phase-matching, one can
vary the laser intensity along the propagation direction to
balance the intrinsic phase with the phase slip due to dispersion.
The required intensity variation to have a constant complex phase
$\arg{\tilde{j_a}}$ in the entire medium is calculated numerically
and shown in Fig. \ref{figure2} for the first interaction zone. It
is optimized for the long trajectory of 50 keV energy but could be
accomplished for any energy value below. Note that only one of the
short and long trajectories can be phase-matched since their
classical actions are different. For the analytical description of
the spatial variations of the laser field in the expression for
$G(x,x^{\prime})$, the eikonal-Volkov approximation is applied
\cite{Avetissian}. This is justified because the additional
driving field causing the modulation 
perturbs the electron energy only slightly.
The experimental realization of the phase-matched scheme  could be achieved, e.g., with a modulated hollow core waveguide.

 We employ a medium length as short as the spatial extent of the APT to minimize
dispersion. In our simulation, each APT consists of 15 pulses with an APT duration of 40 fs.
Our calculations show that in the case of longer APTs, the pulses in the train strongly spread due to dispersion and overlap, thus, violating the condition for the drift compensation.
All pairs of pulses have almost the same coherent contribution to the overall yield. Since the pulses in different zones have experienced a different propagation length through the plasma, their shapes differ slightly. However, phase-matching still can be maintained
by slightly adjusting the modulation profile, as long as the pulse shape still supports the recollision scheme.
The phase-matching scheme imposes a strong demand on the jitter of the laser field $\Delta E/E$:
$\Delta U_p \tau \ll 1$ yields $\Delta E/E\ll (U_p \tau)^{-1}\sim 10^{-4}$.
We choose a gas density of $\rho=10^{19}/$cm$^3$ (ionized by the laser as described before), a diameter of $1$ mm and a length of $12.5\,\mu$m for the interaction volume.
The emitted spectral photon number is shown in Fig. \ref{figure2}.
An integral over the spectrum yields an emitted photon number of
$2.5\times 10^{-8}$ at 50 keV per one collision of APTs which
corresponds to a measureable signal of about 2 photons per day at 1 kHz
repetition rate.
Note that the choice of the atomic species is rather flexible. Multielectron highly charged ions offer an enhanced recombination probability due to core polarization \cite{gordon_kaertner} but produce a larger electron background that can be balanced by a lower gas density. The overall efficiency is maintained or could even be enhanced.
The bandwidth of phase-matched HHG in this scheme is about 150 eV near the cutoff and pulses with a duration of about 35 as can be produced.

\begin{figure}[t]
\begin{center}
 \includegraphics[width=0.23\textwidth]{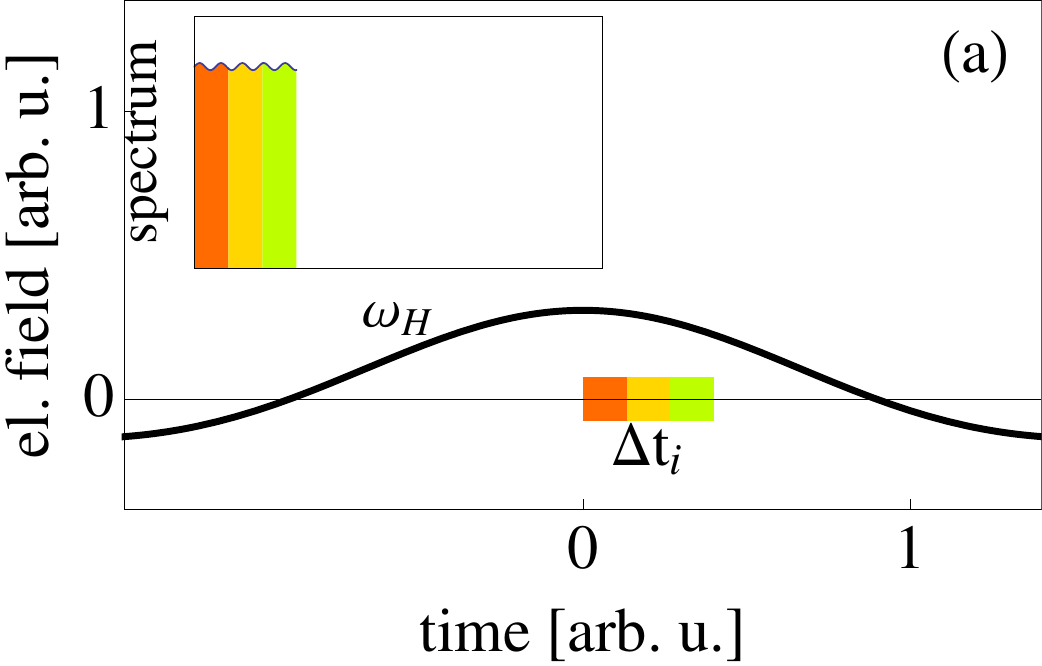}
 \includegraphics[width=0.23\textwidth]{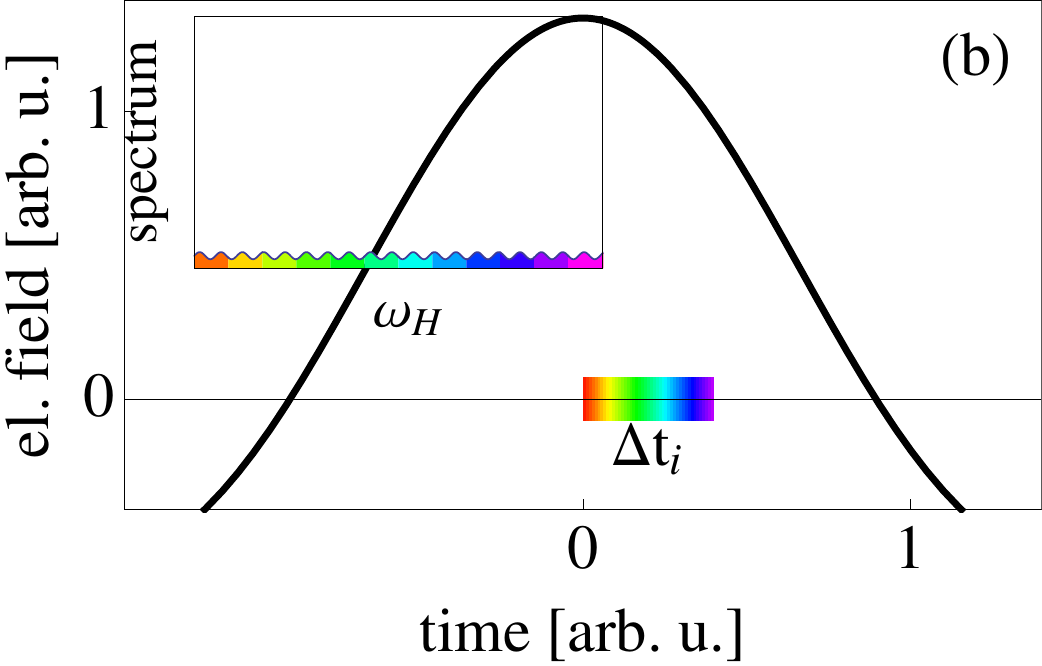}
\caption{Schematic illustration of the chirping factor $ \partial
\omega_H/\partial t_i$: the harmonic bandwidth per unit ionization
time. The insets show harmonic spectra for laser fields (black line) of either low (a) or high (b) intensity. The ionization time window $\Delta t_i$ resulting in HHG is marked on the time-axis, where the colors indicate the harmonic frequency originating from the particular ionization time. 
$\Delta t_i$ remains unchanged under the increase of intensity but the bandwidth of the contained harmonics increases from (a) to (b), consequently, decreasing the ionization propability per harmonic frequency expressed by an increase of the chirping factor.} \label{chirping}
\end{center}
\end{figure}

The small magnitude of the harmonic signal compared to current XUV HHG yields can be explained by investigating the  spectral HHG photon rate $\dot{N_n}$ for phase-matched emission \cite{klaiber_tailored,Ivanov} of the harmonic order $n=\omega_H/\omega$  from a fixed volume
\begin{equation}
\dot{N_n}  \sim w_i(t_i)\left| \langle 0\vert V_H\vert \mathbf{p}\rangle\right|^2 (v_{\perp}^2 \tau^2 \partial\omega_H/\partial t_i)^{-1}.
\label{wn}
\end{equation}
Here $w_i(t_i)$ is the ionization rate with the ionization time $t_i$, $\langle 0\vert V_H\vert \mathbf{p}\rangle$ 
the recombination amplitude and the last factor accounts for the dynamical properties of the wave
packet. $v_{\perp}^2 \tau^2$ expresses the transversal electron spreading with transversal spreading velocity $v_{\perp}$, $\tau$
the excursion time of the electron and $ \partial \omega_H/\partial t_i$ is the so-called electron wave packet chirping factor discussed below. 

We proceed by analyzing the scaling of $\dot{N_n}$ with increasing laser intensity at a harmonic energy near the respective cutoff provided that $w_i(t_i)$ is kept constant by an appropriate choice of $I_p$.
The recombination amplitude decreases with increased electron energy favoring scattering rather than recombination. Its scaling depends on the shape of the ionic potential: $\vert \langle 0\vert V_H\vert\mathbf{p}\rangle_{C}\vert^2\sim I_p^{5/2}/\omega_H^4$ for a
hydrogen-like ion  and $\vert \langle 0\vert V_H\vert\mathbf{p}\rangle_{Z}\vert^2\sim \sqrt{I_p}/\omega_H^2$
for a zero-range potential  with $I_p\ll\omega_H$ and $p^2 \sim \omega_H$.
Regarding the last term of Eq. \eqref{wn},  we follow  \cite{Ivanov} to find $v_{\perp}=\sqrt{E}/I_p^{1/4} \sim(\omega_H/I_p)^{1/4}$ and illustrate
 the chirping factor in Fig. \ref{chirping}. It describes that the bandwidth of the harmonics emitted from a fixed ionization time window rises with increasing laser intensity (i.e. the ionization probability per harmonic decreases) and  can be estimated as $\partial \omega_H/\partial  t_i\sim\omega_H /{\Delta t_i}$.
Thus, the photon emission rate in a constant bandwidth for a zero-range potential scales as $\dot{N_n}\sim  I_p/\omega_H^{3.5}$. A rough estimate for the scaling of $I_p$ at a constant ionization rate can be derived fixing the common  tunneling exponent yielding $I_p\sim E^{2/3}\sim \omega_H^{1/3}$ and, consequently, $\dot{N_n} \sim 1/\omega_H^{3.17}$. The decrease for a hydrogen-like potential is even more dramatic. Therefore, the HHG photon yield decreases with rising photon energy due to the decreased probabilities of ionization per harmonic and the reduced recombination cross section. Our analysis points out a possible future direction for optimization of HHG by means of increasing the ionization time window at a given harmonic bandwidth.

In conclusion, we have shown that the drift and phase-matching do
not restrict HHG to the non-relativistic regime. The proposed
setup renders the relativistic regime of HHG in a multi-atom ensemble accessible.

\end{document}